\documentclass{article}
\usepackage{blindtext}
\usepackage[utf8]{inputenc}
\usepackage{graphicx}
\usepackage{enumerate}
\usepackage{amsmath}
\usepackage{amsthm}
\usepackage{amsfonts}
\usepackage{mathtools}
\usepackage{indentfirst}
\usepackage{listings}
\usepackage{xcolor}
\usepackage{subfig}
\usepackage[margin=1in]{geometry}
\usepackage{enumitem}
\usepackage{lineno}
\usepackage{float}
\usepackage{cite}
\title{The number of immune defense and counter-defense systems sustained in the arms race between procaryotes and viruses}
\author{Yaroslav Ispolatov\footnote{Department of Physics, Center for Interdisciplinary Research in Astrophysics and Space Science, University of Santiago of Chile} , Anna Lekontseva\footnote{Moscow Institute of Physics and Technology, 9 Institutskiy per., Dolgoprudny, Moscow Region, 141701, Russia}, and Konstantin Severinov\footnote{Waksman Institute of Microbiology, Rutgers, The State University of	New Jersey, Piscataway, NJ, USA}}

\date{\today}

\begin{document}

\maketitle
\begin{abstract}
Prokaryotes have evolved various mechanisms to counter viruses, which in their turn developed numerous strategies to avoid defenses of the hosts. Dozens of such defense and counter-defense mechanisms have recently been discovered, yet the number of such systems held by a given virus or its host is limited. Here, we present numerical and theoretical arguments for the existence of the maximal number of ecologically and evolutionary sustainable defense and counter-defense systems maintained by both sides at any time of the never-ending evolutionary arms race. We find that the number of such systems is of the order of 10 for a broad range of assumptions about the costs and benefits of defense and counter-defense mechanisms and their specificity. This optimum in the number of defense and counter-defense systems appears as a result of a compromise between the metabolic and autoimmune costs of adding a new layer of defense and the benefits it conveys.      
\end{abstract}

\section*{Significance statement}
Almost all living creatures are either predators or prey (and many are both). At all scales of life, from single cells to largest mammals, prey evolve defenses against predators and predators in their turn evolve counter-defense mechanisms. As a result of this ever-lasting arms race, each organism usually possesses multilayer defense and counter-defense arsenals. Is there a limit to the size of these arsenals? In this work we present a proof-of- principle argument that such a limit indeed exists. Using for definity an example of prokaryotes as the prey and their viruses as predators, we show that the number of distinct defense and counter-defense system that the prey and predators possess is limited by a number of the order of 10 and is only weakly dependent on the ecology of predator-prey interactions.  Similar principles and limits likely emerge universally in various antagonistic conflicts such as crime and law enforcement or real arms race. 

\section{Introduction}
Prokaryotes are involved in a constant conflict with mobile genetic elements, most prominently, with viruses (phages). Individual encounters between the parties of the conflict often result in elimination of either or both sides. 
The conflict, lasting for billions of years, is driven by an arms race with hosts evolving immune defenses to protect themselves against the viruses, while the latter developing counter-defenses to bypass the host defences. As a result, a typical prokaryotic cell (in the following referred to as a "cell" for brevity) possesses a wide repertoire of defense systems that, among others, prevent viral adsorption, degrade
foreign nucleic acids injected inside the cell, deplete infected cells of essential molecules, induce  dormancy and death of infected cells, or synthesize molecules that inhibit viral replication 
\cite{hampton2020arms, agapov2024multi, millman2022expanded, wang2023unveil}.
Viruses, in their turn, have evolved various counter-defense strategies that allow them to bypass host cells defenses \cite{safari2020interaction}. The diversity of available defense and counter-defense systems requires a certain level of redundancy both in the immune arsenal a cell possesses and in the collection of viral counter-defense strategies. Thus, it is interesting to examine how many layers of immune defense systems and counter-immune strategies cells and viruses can evolve in total? And how many of such strategies does a given cell or virus  typically possess? Recent estimates of the number of mechanistically distinct defense systems that a typical prokaryotic cell carries is between 5 and 6  \cite{agapov2024multi, millman2022expanded}, which is an order of magnitude less than the total number of such systems discovered to date, with many more systems remaining to be discovered. Furthermore, the experimental evidence suggests that the diversity  of counter-defense systems constrains the arms race dynamics, making it less probable or even preventing the evolution of their reciprocal defense systems \cite{castledine2022greater}.  These characteristics of defense and counter-defense arsenals  are essential to understanding the ecology of coexistence of prokaryotes and their viruses that essentially shapes the Earth's biosphere.

The existing theoretical and numerical analysis of population dynamics of cells and phages are mostly focused on explaining sustainability of their coexistence by the emergent spatial structures \cite{eriksen2020sustainability}, diversification of viral strategies \cite {kimchi2024lytic}, and synergistic effects of various defense systems of cells \cite{arias2022coordination, wu2024bacterial}. To the best of our knowledge, only the recent work \cite{kimchi2024bacterial} (posted on bioRxiv after the main part of our analysis was completed) analyzes the
effect of various layers of defense and counter-defense systems on bacterial and viral population dynamics. 

Here, we attempt to provide conceptual arguments for the existence of an upper limit in the number of such layers. Under a broad range of assumptions, such limits appear as a result of a compromise between the benefits provided by adding new defense mechanisms and their metabolic and self-immune costs. These limits exist both for cells, where at some point an addition of a new level of defense becomes detrimental, and for viruses, where the cost of a new evasion strategy exceeds its benefit. Using a generic Lotka-Volterra description for population dynamics of interacting cells and viruses, we show that these limits universally exist for various dependencies of the costs on the number of defenses and counter-defense strategies.   
We first present simulations where defense and counter-defense mechanisms are nonspecific. This example reveals the upper limit on the number of defense system a cell can evolve under given ecological conditions and metabolic costs of these systems.  It is followed by a model that describes more common settings where each defense system can be counteracted only by a specific counter-defense. The defense and counter-defense systems are assumed to be available from a commonly accessible pool via, e.g., horizontal gene transfer or random mutation. We compare the maximal number of sustainable defense layers in both models and show how variations in costs, such as metabolic load and reduction in attack rates, affect the ecological sustainability of these mechanisms. Finally, we offer a theoretical estimate of the maximal number of immune layers, which almost perfectly agrees with our simulation results.

\section {Model}
\subsection{Population dynamics}
Consider a community populated by strains of cells and viruses, distinct only in the repertoires of
their defense and counter-defense systems. The community is well-mixed and has no spatial structure, so at each time point its ecological state is completely characterized by population densities of cellular and viral strains  $B_i(t)$  and $P_j(t)$. Different strains of cells are labeled by the number of their immune defense systems $i=0,1,\ldots$, while the label $j=0,1,\ldots$ of a viral strain is the number of defense evasion mechanisms it encodes. In the following, the defense systems and defense evasion mechanisms are also referred to as defense and counter-defense layers. Later, the labels of cellular and viral strains will also indicate the specific types of defense and counter-defense systems those strains possess.

The dynamics of populations $B_i(t)$  and $P_j(t)$ are assumed to be given by a Lotka-Volterra system of equations,
\begin{align}
  \label{LV}
  \frac{1}{B_i}\frac{dB_i}{dt}=\beta_i - \frac{\sum_s B_s}{K} - \sum_j \alpha_{ij}P_j, \;\; i=0,1,\ldots;\\
  \nonumber
  \frac{1}{P_j}\frac{dP_j}{dt}=\chi_j \sum_i \alpha_{ij}B_i - \delta, \;\; j=0,1,\ldots .
  \end{align}
  Here $\beta_i$ is a per capita reproduction rate of cells that depends on the number of immune defense layers $i$, $K$ is the environmental carrying capacity, and $\alpha_{ij}$ is the attack rate at which the viral strain $j$ infects the cellular strain $i$. The attack rate depends not only on the number of immune defense layers $i$ but also on the number $j$ of counter-defense mechanisms of the virus. The summation runs over all viral strains $j$, which reflects that in principle any of those strains can attack the cellular strain $i$. In the second line, the per capita reproduction rate of the virus includes the factor $\chi_j$, which comprises the burst size and the rate at which the viral offspring  are produced during a successful infection. This rate, and consequently $\chi_j$, decreases with the number $j$ of defense evasion mechanisms of the virus due to their metabolic cost. All viruses have an identical per capita death rate $\delta$.

  The main feature of our model are the dependencies of $\beta_i$, $\alpha_{ij}$, and $\chi_j$ on $i$ and $j$.

\subsection{Cellular growth rate}
  The metabolic and autoimmune cost of maintaining a certain number of immune defense layers results in a slower growth. Hence, $\beta_i$ is a decreasing function of $i$. There are two simple possibilities to define this decrease. First is additive, so that each new layer slows down the growth rate by a constant, 
  \begin{align}
  \label{ba}
  \beta_i=\beta_0(1-iC_{\beta}), 
  \end{align}
  with $\beta_0$ and $C_{\beta}$ being parameters. Because the reproduction rate has to be positive, the choice of this form of $\beta_i$ artificially sets a limit on the maximal number of immune defense layers, $i<1/C_{\beta}$. Hence, in  all meaningful scenarios $i$ must stay below this limit.

Another possibility is to define the growth rate penalty multiplicatively: each new layer decreases the birth rate by a constant multiplicative factor
 \begin{align}
  \label{bm}
\beta_i=\beta_0(1-C_{\beta})^i. 
 \end{align}
This form in principle allows a cellular strain to evolve unlimited number of defense layers. 

We will analyze Eq. (\ref{LV}) using both these forms. 

\subsection{Viral growth rate}
Similarly to the case of cells, an addition of a immune avoidance mechanism slows viral growth. So $\chi_j$ must also be a decreasing function of $j$, and similarly to cellular growth rate, it could be defined both in additive and multiplicative ways, 
  \begin{align}
  \label{ha}
  \chi_j=\chi_0(1-jC_{\chi}), 
  \end{align}
or 
  \begin{align}
  \label{hm}
  \chi_j=\chi_0(1-C_{\chi})^j. 
  \end{align}

%We will consider both these forms. 

\subsection{The attack rate}

Unlike the two previous cases, the attack rate (or the level of protection by an immune mechanism) is always multiplicative: it is well-established that a successful immune defense system reduces the number of infected cells by a certain factor (e.g., measured as a "protection level" in \cite{kirillov2022cells}). We assume that different immune protection systems act independently, so that the protection level offered by several independent systems is the product of protection levels offered by each system separately. We also assume that each immune avoidance mechanism works perfectly against its immune system and allows a virus to restore the attack rate to the pre-immune level. This suggests the following form for the attack rate,
 \begin{align}
  \label{ar}
  \alpha_{ij}=\alpha_0(1-C_{\alpha})^{(i-j)\theta(i-j)}, 
  \end{align}
where $\theta(i-j)$ is a step function, 
 \begin{align}
  \label{th}
 \theta(x)={\begin{cases}1&{\textrm{if }}x\geq 0\\0&{\text{if }}x<0\\\end{cases}}
   \end{align}
In other words, (\ref{ar}) means that a virus does not gain any increase in attack rate if it evolves an avoidance mechanism against the immune system that the cellular strain it attacks does not possess. It is also assumed in (\ref{ar}) that the immune systems and avoidance mechanisms are non-specific, so the net effect of $i$ immune layers and $j$ avoidance mechanisms depends only on  ${(i-j)\theta(i-j)}$, but not on specific pairings between them. A sketch that illustrates this scenario is presented in the left panel of Supplementary Figure 1.
\subsection {Simulations}

We perform two types of simulations:

In an ecological-only variant, all cellular and viral strains are seeded at a small initial density, 
\begin{align}
    \label{eco}
    B_i(t=0)=B_{seed},\; P_j(t=0)=P_{seed}\;\; \forall \{i,j\},
\end{align}
and are allowed to evolve according to Eqs. (\ref{LV}) in a totally deterministic way.

In the second variant that mimics evolution, the simulation is initiated with small densities of cell without immune defenses and viruses without avoidance mechanisms,
\begin{align}
    \label{evo}
B_0(t=0)=B_{seed}, P_0(t=0)=P_{seed}, \;\textrm {and }B_i(t=0)=P_j(t=0)=0 \;\;\forall \{i,j\}\geq 1.
\end{align}
With the frequency proportional to its density, a given strain produces a new (mutant) strain that either loses or gains one  defense (or counter-defense) system. %The interval of time between successive mutational events is set proportional to the sum of all population densities. 
Specifically, an event occurs when the time integral of the sum of all population densities reaches a certain threshold (usually set equal to 500) and then an ancestral strain is chosen with the probability proportional to its population. Mutational events that produce already existing strains are ignored as their contributions to already existing  populations are negligible.  Strains with population below a certain limit (equal to $B_{seed}/2$ or $P_{seed}/2$) are declared extinct and their densities are set to zero. 

The parameters used in first set of simulations are shown in Table \ref{tab:table1}. 
\begin{table}[h!]
  \begin{center}
    \caption{Values of parameters}
    \label{tab:table1}
    \begin{tabular}{|l|c|r|} 
            \hline  
      \textbf{Parameter} & \textbf{Value} & \textbf{Meaning}\\
      \hline
$\beta_0$ & 10 & Cellular birth rate amplitude\\
     \hline
$C_{\beta}$ & 0.1 & Cellular birth coefficient\\
     \hline
$K$ & 1 & Environmental carrying capacity\\
      \hline  
$\alpha_0 $ & 0.1 & Attack rate amplitude\\
     \hline
$C_{\alpha}$ & 0.3  & Attack rate coefficient \\
      \hline  
$\chi_0$ & 5 & Viral birth rate amplitude\\
      \hline
$C_{\chi}$ & 0.1 & Viral birth rate coefficient\\
      \hline  
$\delta$ & 1 & Viral death rate\\
      \hline  
$B_{seed}$ & $2\times 10^{-6}$ & Initial density of cells\\
      \hline  
$P_{seed}$ & $2\times 10^{-6}$ & Initial viral density\\
      \hline  
    \end{tabular}
  \end{center}
\end{table}
In the subsequent examples we will show the quantitative effects of varying $C_{\alpha}$, $C_{\beta}$ and $C_{\chi}$.
The phenomenology that we describe below is very robust and is reproducible for a wide range of parameters. 
\section{Results}
\begin{figure}[h!]
    \centering
  \includegraphics[width=.98\linewidth]{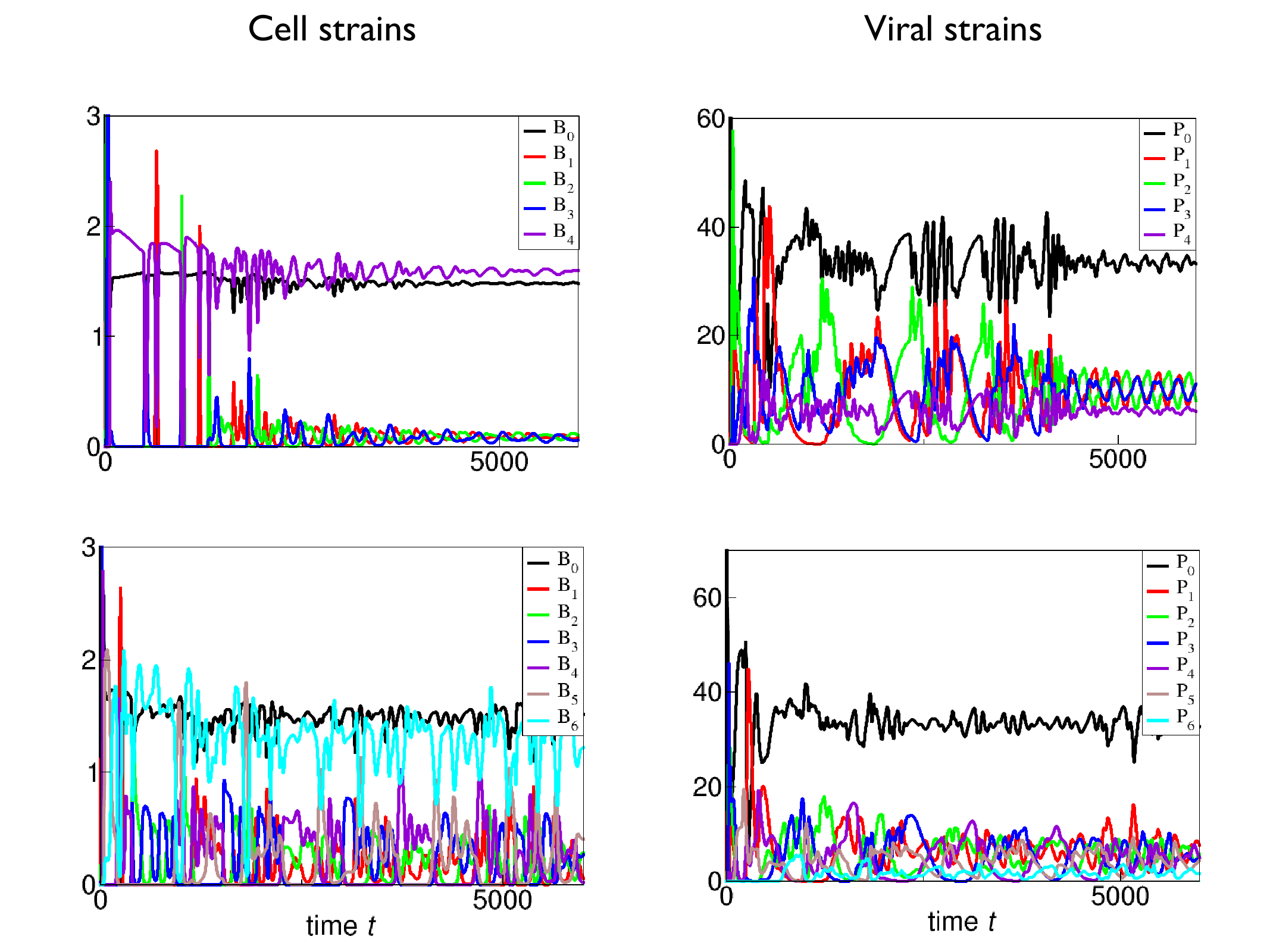} 
    \caption{The figure shows the population dynamics of surviving cellular (left panels) and viral (right panels) strains. The scenarios with  additive (given by Eqs. (\ref{ba},\ref{ha})) 
    and multiplicative  (given by Eqs. (\ref{bm},\ref{hm})) metabolic costs for cells and viruses  are shown in the first and second rows.  
    The simulation is initiated with small population densities of all cellular and viral strains with no subsequent mutation or extinction events and all parameters are taken from  Table \ref{tab:table1}.}
    \label{f1}
\end{figure}

We start with ecological simulations (\ref{eco}) and first examine all four possible combinations of multiplicative and additive costs for viruses and cells. Our intention is to show that the existence of a limit on the number of defense and counter-defense layers is universal. Indeed, for the parameters listed in Table 1, the additive costs for the cells (\ref{ba}) and viruses (\ref{ha}) limit the number of layers to 4, top row in Fig. \ref{f1}, 
while the less steep multiplicative costs (\ref{bm},\ref{hm}) allow the evolution of 6 such layers, bottom row in Fig. \ref{f1}. 

In "mixed" scenarios with multiplicative costs for cells and additive for viruses and vice versa, shown in Supplementary Fig. 2, 5 layers of defense and counter-defense are sustained. 

\begin{figure}[h!]
    \centering
 \includegraphics[width=.98\linewidth]{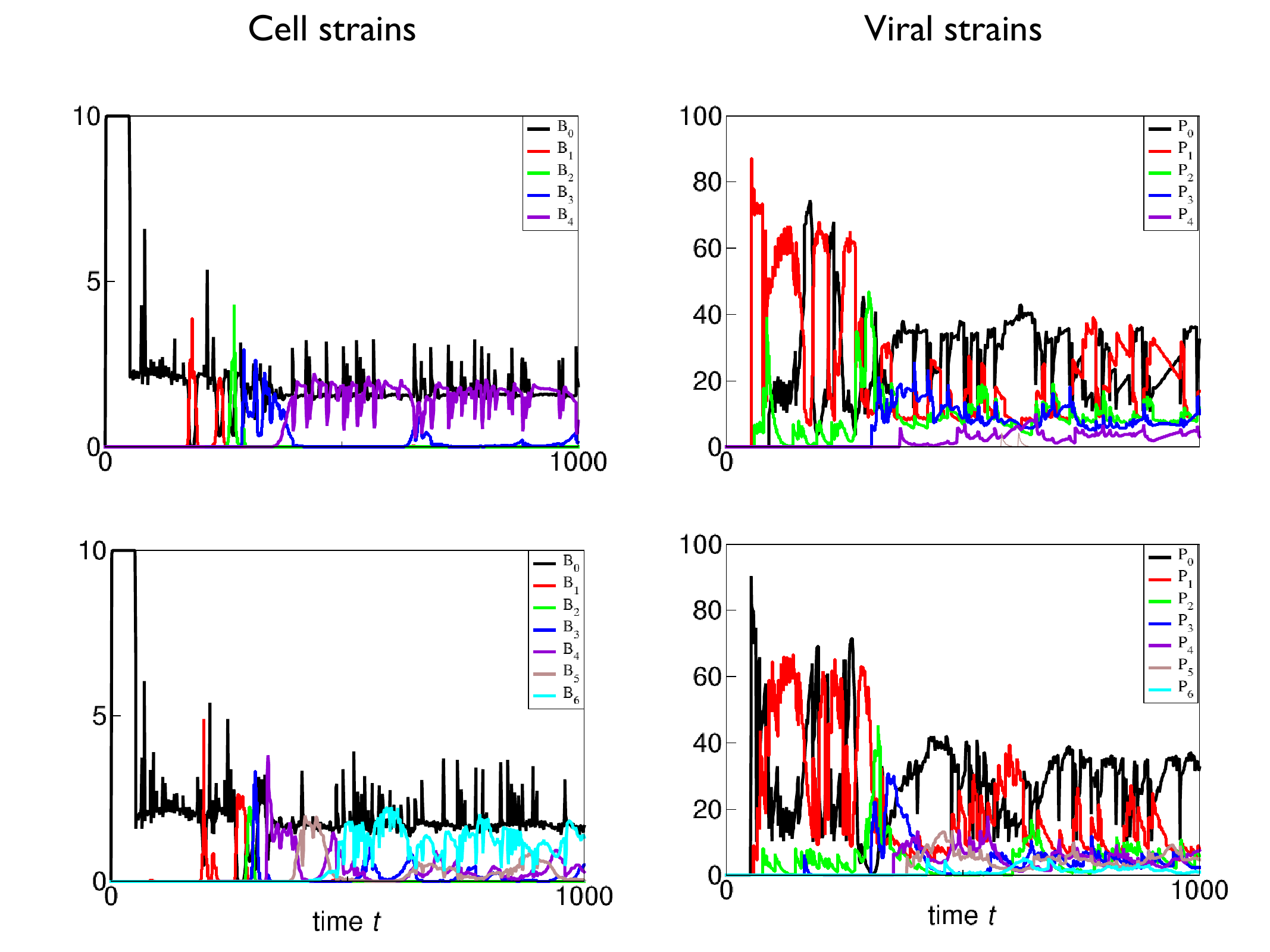} 
    \caption{Same as in Fig. \ref{f1}, but with the evolutionary simulation instead of the ecological one.  Note that the cellular strains with the intermediate number of levels of defense become extinct at later time.}
    \label{f5}
\end{figure}

Simulations with evolutionary scenario defined by Eqs. (\ref{evo}) and illustrated in  Fig \ref{f5} predict the same maximal number of levels of immune defenses and avoidance mechanisms as their ecological counterparts. Yet the evolutionary simulations may also result in the extinction of  strains of cells with intermediate numbers of immune layers (such as $B_1$ and $B_2$ in Fig. \ref{f5}).

The generally smaller number of sustainable layers observed for the additive costs is not surprising as the additive costs increase faster with the number of layers than the multiplicative ones, 

\begin{align}
    \label{ma}
    {1-iC}<(1-C)^i \;\textrm{for}\; C<1 \;\textrm{and}\; i\geq2
\end{align}

Next we show how the variation of costs $C_{\beta}$ and $C_{\chi}$ affects the maximal number of layers. As expected, an increase in costs from 0.1 to 0.15  
reduces the number of defense layers and avoidance mechanisms with additive costs to 2  (first row in Fig. \ref{f7}), and with multiplicative costs  to 3 (first row in Supplementary Fig. 3.

\begin{figure}[h!]
    \centering
    \includegraphics[width=.98\linewidth]{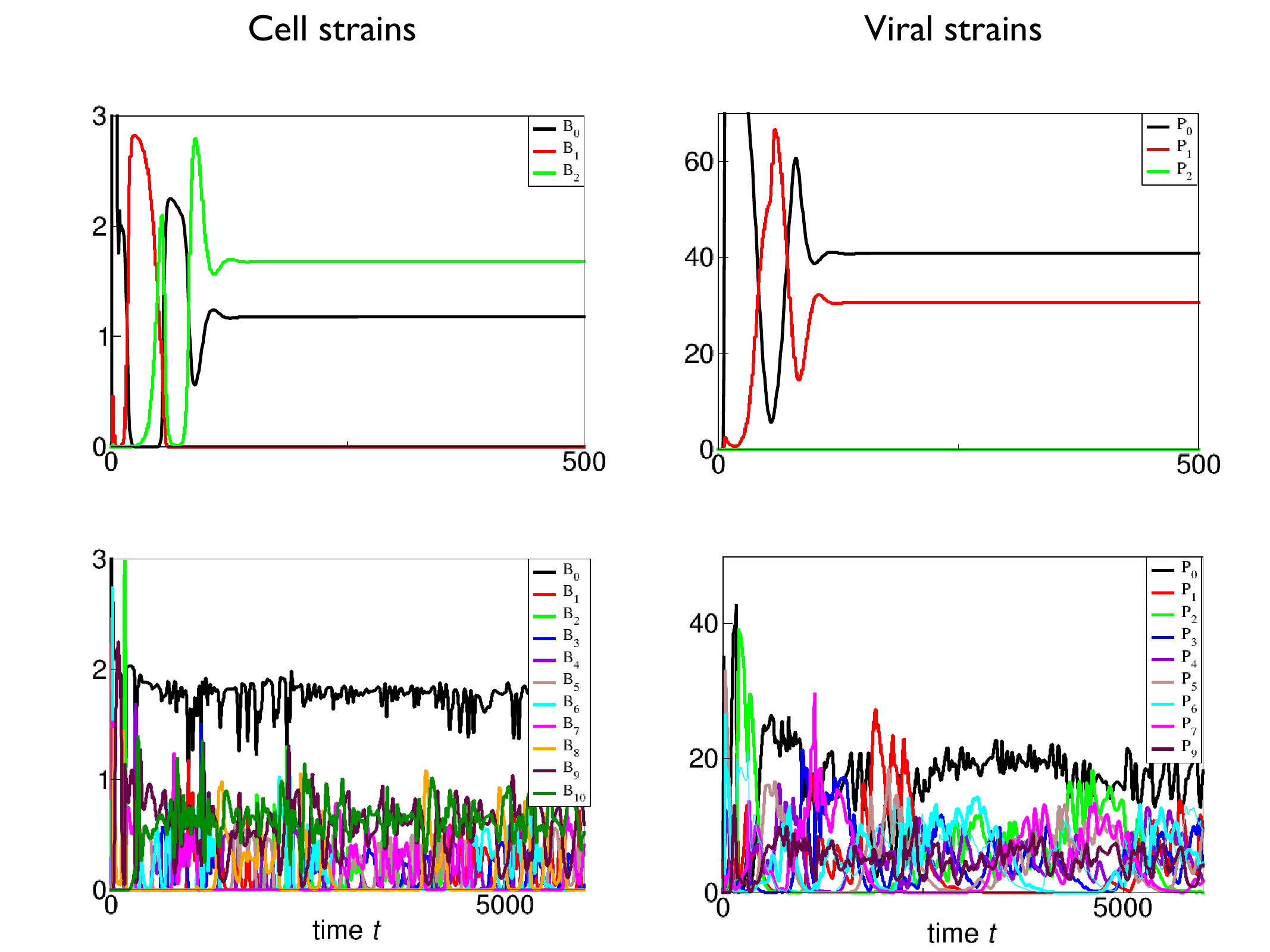} 
    \caption{Same as in first row of Fig. \ref{f1}, but with the additive metabolic costs for cells and for viruses being increased, $C_{\beta}=C_{\chi}=0.15$, first row, and decreased, $C_{\beta}=C_{\chi}=0.05$, second row.}
    \label{f7}
\end{figure}

Reciprocally, a reduction in costs from 0.1 to 0.05 results in a noticeable increase in the number of sustainable immune and counter-immune systems. In second row in Fig. \ref{f7}, the population dynamics of 11 cellular strains sustaining up to 10 level of immune defense and 10 viral strains with up to 9 avoidance mechanisms are shown.

An increase in $C_{\alpha}$ from 0.3 to 0.5, which translates into a greater reduction in attack rate with each additional layer, also results in an increase in the number of layers even if the cost of the layers in maintained high (equal to 0.15). Supplementary Fig. 4 shows that such an increase allows cells to sustain up to 4 layers of immune defense, up from 3 sustainable for $C_{\alpha}=0.3$. However, viruses were able to sustain only 3 layers of avoidance mechanisms, similarly to the reference case with $C_{\alpha}=0.3$.

When the metabolic cost of an immune system is less than the cost of an avoidance mechanism, an extra layer of protection above the previous maximal number of immune defense strategies may evolve, Supplementary Fig. 5. However, in reverse scenario when the cost for cells is higher than that for viruses, fewer layers are sustained, Supplementary Fig. 6.

\section{An estimate for the maximal number of layers}

A simple estimate allows us to quite accurately predict the maximal number of sustainable defense layers. It is based on the adaptive dynamics derivation (see, for example, \cite {dieckmann1996dynamical}, adjusted to the discrete nature of the main evolutionary variables, the number of layers $i$ and $j$. 

Consider a resident cellular strain with $i$ levels of defense coexisting with a resident viral strain with the same number ($j=i$) of avoidance mechanisms. The resident strains are assumed to be in an ecological equilibrium or at steady state population levels. As it follows from the figures above, the population dynamics does necessarily converge to a steady state. Nevertheless, in the idealized case of one cell and one viral strain, it is well known that the Lotka-Volterra population dynamics converges to the steady state for any finite carrying capacity $K$. 
The steady population densities $\bar B_i$ $\bar P_i$ are given by the solution of the system of equations obtained from (\ref{LV}) by setting the left-hand sides equal to zero,
 \begin{align}
  \label{st1}
 {\begin{cases} \beta_i - \frac{\bar B_i}{K} - \alpha_{0}\bar P_i=0&\\
 \\
 \chi_i \alpha_{0}\bar B_i - \delta=0&\\\end{cases}}.
   \end{align}
The solution reads
 \begin{align}
  \label{st2}
 {\begin{cases} \bar B_i=\frac{\delta}{\alpha_0 \chi_i}&\\
 \\
 \bar P_i=\frac{\beta_i-\delta/(\alpha_0\chi_i K)}{\alpha_0}&\\\end{cases}}.
   \end{align}
Now consider the per capita growth rate of a rare cell mutant with $i+1$ layers of immune defense in the presence of resident cell and viral strains $B_i$ and $P_i$. When the growth rate is negative, the strain $B_i$ will not evolve more defense layers, so $i$ becomes the maximal number of sustainable immune defense layers. (Note that the rare viral mutant $P_{i+1}$ will always have a negative growth rate as its reproduction rate will be less than that of $P_i$ by a factor $\chi_{i+1}/\chi_i$, and the reproduction rate of $P_i$ equals its death rate at the steady state.)
 \begin{align}
  \label{imax1}
\beta_{i+1}-\frac{\bar B_i}{K} - \alpha_{1}\bar P_i<0
   \end{align}
Substitution of (\ref{st2}) in (\ref{imax1}) yields
\begin{align}
\label{imax2}
\beta_{i+1}-(1-C_{\alpha})\beta_i-\frac{\delta C_{\alpha}}{\alpha_0\chi_iK}<0
\end{align}
For the multiplicative costs (\ref{bm},\ref{hm}) the solution of ({\ref{imax2}}) is particularly simple,
\begin{align}
\label{imax3}
[(1-C_{\beta})(1-C_{\chi})]^i<\frac{\delta C_{\alpha}}{\alpha_0\chi_0\beta_0 K (C_{\alpha}-C_{\beta})},
\end{align}
which yields the estimate $I_{max}$ for the maximal number of sustainable immune defense layers, 

\begin{align}
\label{imf}
I_{max}=\left.{\ln\left[ \frac{\delta C_{\alpha}}{\alpha_0\chi_0\beta_0 K (C_{\alpha}-C_{\beta})} \right]}\right/{
\ln\left[ (1-C_{\beta})(1-C_{\chi})\right]}.
\end{align}

This simple estimate works surprisingly well and produces values of $I_{max}$ that round up to the observed values ($i_{max}$) in all cases that we have considered,  see Table \ref{tab:table2}.
\begin{table}[h!]
  \begin{center}
    \caption{Estimated ($I_{max}$) and observed ($i_{max}$) maximal number of immune defense layers. Parameters that are not listed in column 3 are the same as in Table \ref{tab:table1}.}
    \label{tab:table2}
    \begin{tabular}{|l|c|r|} 
            \hline  
      $\mathbf{I_{max}}$ & $\mathbf{i_{max}}$ & \textbf{Parameters}\\
      \hline
5.7 & 6 & Table \ref{tab:table1}\\
     \hline
2.8 & 3 & $C_{\beta}=C_{\chi}=0.15$\\
     \hline
13.9 & 14 & $C_{\beta}=C_{\chi}=0.05$\\
      \hline  
3.9 & 4& $C_{\beta}=C_{\chi}=0.15,\;C_{\alpha}=0.5$\\
     \hline
    \end{tabular}
  \end{center}
\end{table}
%The estimate 

\section{Specific defense and counter-defense mechanisms}
\label{spec}
\subsection{Definition of specific immunity and counter-immune defense}
So far, we have analyzed the scenario where each defense level offered universal protection, and each counter-defense mechanism worked against any defense system. This simplistic approach sets the upper limit on the maximal number of defense and counter-defense layers: each additional defense mechanism protects the cell strain against all viruses, and each additional counter-defense layer allows a virus to overcome all existing defense mechanisms.  We expect that the specificity of defense and counter-defense mechanisms will decrease the sustainable number of these layers. This is so because an addition of a new defense system does not necessarily protect against all viral strains with their specific counter-defenses and thus brings less benefits than in the universal case, while the costs are identical. Here we  demonstrate it with a  model where only a specific counter-defense mechanism can overcome a particular immune defense. 

Assume that there exists a pool of $N$ distinct immune defense systems and the same number of viral counter-defense mechanisms each specific to its “cognate” defense system. A cellular strain is now labeled with several subscripts that describe its repertoire of immune defenses. The number of such subscripts can be any integer from zero to $N$. For example, the population $B_{1,2,4}$ has $m=3$ immune systems, the first, the second, and the fourth. Similarly, viral strains are labeled by their counter-defense mechanisms: for example, the population $P_{2,3}$ has two such mechanisms, the second and the third. When a cell has an immune system and the virus that attacks it does not possess the corresponding counter-defense, the attack rate is reduced by the factor $1-C_{\alpha}$. This rule replaces the previously assumed form of the attack rate given by Eq. (\ref{ar}). So in our example the attack rate of virus $P_{2,3}$ on the cell $B_{1,2,4}$ is $\alpha_0 (1-C_{\alpha})^2$ since the virus does not have counter-defense against the first and the fourth immune systems.  A sketch of the specific defense -- counter-defense scenario is presented in the right panel of Supplementary Figure 1. The additive metabolic costs of each immune defense and counter-defense layer are assumed to be independent of the nature of these systems and are given by Eqs. (\ref{ba}, \ref{ha}) as before.   

%As in the previous example,  we run two types of simulations. In the ecological simulations, populations with all possible combinations of defenses and counter-defenses are present at small densities from the beginning. The evolutionary simulations start from cell and viral strains without any defense and counter-defense mechanisms, correspondingly. These mechanisms are subsequently acquired via mutational events that occur in the same way as described in Section 2.5: In each mutational event a new cell (or viral) strain is formed with one more or one less defense (counter-defense) mechanism than the ancestral strain.  
In this setting the ecological simulation appears more relevant: the cellular and viral strains with all possible combinations of defenses and counter-defenses are present at small densities from the beginning. Subsequent population dynamics of the strains are defined by Eqs. (\ref{LV}).

\begin{figure}
    \centering
        \includegraphics[width=.98\linewidth]{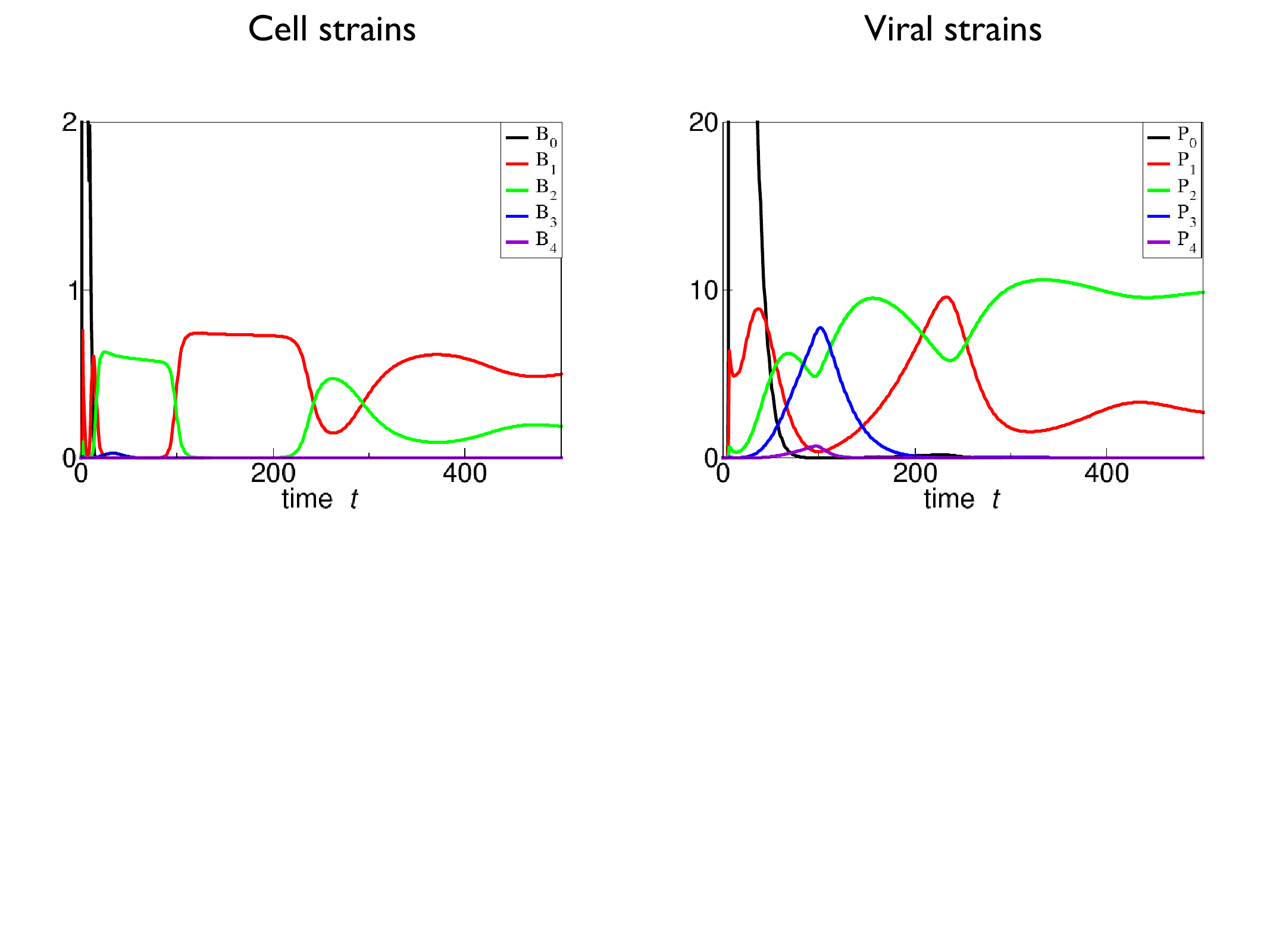} 
    \caption{ The population dynamics of cellular (left panel) and viral  (right panel) strains calculated with the specificity of defense and counter-defense mechanisms taken into account (as described in Section \ref{spec}). Strains with the same number of layers have the same populations densities and are shown by single lines.  The metabolic costs for cells and for viruses are additive and given by Eqs. (\ref{ba},\ref{ha}) and the parameters are listed in Table \ref{tab:table1}.
    The simulation is initiated with small population densities of all cellular and viral strains with no subsequent mutation or extinction events. The figure reveals that even though strains with 3 and 4 layers at a certain time develop non-vanishing population density, in the long term only 2 levels of specific immune defense or counter-defense mechanisms are sustainable.}
    \label{f_spec_ecol}
\end{figure}

The simulations confirm our qualitative prediction that taking into account the specificity of immune defenses and counter-defenses reduces the maximal number of such layers that are sustained by cells and viruses. Fig. \ref{f_spec_ecol} shows  the results for the maximal number of defense and counter-defense layers $N=4$ and other parameters taken from the Table 1. It follows that surviving strains have at most 2 defense layers, while in the non-specific simulations the number of both defense and counter-defense layers was 4, Fig. \ref{f1}.

\subsection{Combinatorial arguments}
Here we explain why the specificity of immune defenses and counter-defenses reduces the maximal number of such layers compared to the simplified scenario of universal layers. Consider the particular example of  $N=4$  distinct defense and counter-defense systems and weigh the benefits and costs for a cell to acquire the third defense mechanism on top of the already existing two, $m=2$. We assume that 
$\binom{N}{m}=6$ strains of cells, each possessing a distinct combination of two defense layers, coexist in the system. Because all defense mechanisms have the same efficiency and metabolic costs, the population densities of these 6 strains are the same. We also assume that cells of each $B_{pq}$ strain have counterpart viruses of the $P_{pq}$ strain with the corresponding counter-defense mechanisms. Populations densities of these 6 viral strains are also equal as their ecological parameters are identical. The attack rate on a cell of any resident strain originating from all 6 viral strains is 
\begin{align*}
 A_{resident}=\frac{\alpha P}{6}\left[1 + 4(1-C_{\alpha}) + (1-2C_{\alpha})\right],   
\end{align*}
where $P$ is the net population of all viral strains. The first term in the square brackets originates
from attacks of the viral strain with the counter-defense mechanisms perfectly matching the set of defenses, the second term describes the attacks from $N=4$ viral strains with one mismatch between the defense and counter-defense mechanisms, while the third term accounts for the viral strain with the mismatches in both counter-defenses. 
Now assume that one of the cellular strains, let it be $B_{1,2}$ without loss of generality, produces a mutant $B_{1,2,3}$, acquiring the third defense mechanism. The attack rate on the mutant cell from all resident viral strains is   \begin{align*}
 A_{mutant}=\frac{\alpha P}{6}\left[3(1-C_{\alpha}) + 3(1-2C_{\alpha})\right],   
\end{align*}
Three first terms result from the attacks from viruses $P_{1,2}, P_{2,3},$ and $P_{1,3}$ that have to overcome one defense layer, while the last term describes the attacks from viruses $P_{1,4}, P_{2,4},$ and $P_{3,4}$ that have to overcome two defense layers. The mutation brings in the reduction in the attack rate
\begin{align*}
 \Delta A
 = A_{resident} -  A_{mutant}=-\frac{\alpha C_{\alpha} P}{2}.
\end{align*}
Yet the reduction in the attack rate in the simple model (Sec. 2) with non-specific defense and counter-defense mechanism is twice larger,  
\begin{align*}
 \Delta A_{non-specific} =-{\alpha C_{\alpha} P}.
\end{align*}
At the same time, the costs of an additional defense systems are identical in both model and are expressed as a decrease in the birth rate given by (\ref{ba}). This explains why the more realistic consideration that takes into account the differences between various immune defense systems and specificity of each counter-defense mechanism predicts fewer sustainable defense and counter-defense layers.  

\section{Discussion}
 We presented a proof of the principle  that the numbers of defense systems of any cell  and counter-defense strategies of any virus are finite: an addition of a new defense or counter-defense system above a certain threshold carries the metabolic and self-immune costs that are higher than the benefits it conveys. 
 
 There are two main conclusions. In the first part of our study that mimics innate immunity, where the defense and counter-defense systems are non-specific, an addition of a new defense system makes the cell strain immune against all existing viral strains (until one of them evolves the counter-defense strategy). Thus, it estimates the maximal number of defense systems that are evolvable under given ecological parameters and metabolic costs of defenses. In the second part of our study where specificity is introduced, such that each counter-defense allows a virus to overcome only a particular defense system,  the number of defense and counter-defense layers sustainable in the long term is less than in the non-specific case. This model mimics the scenario of acquisition of defense and counter-defense systems by adding spacers to a CRISPR array and explains the observation that a cell employs significantly fewer defense systems that are known to exist, \cite{millman2022expanded, wang2023unveil}.   

We also found that both ecological and evolutionary simulations predict similar limits on the number of defense and counter-defense layers, with the latter often exhibiting extinction of strains with intermediate numbers of defense layers. Furthermore, variations in metabolic costs play a critical role in determining the maximal number of layers. Predictably, a lower cost leads to an increase in number of sustainable layers, while a higher cost limits the system to fewer defense and counter-defense systems.

A key numerical takeaway from our findings is that the number of sustainable layers of immune defense or counter-defense systems is of the order of 10. The analytic estimate we provide for the maximal number of sustainable layers, which is confirmed by our simulations, offers a simple explanation for this limit: The maximal number of sustainable levels depends only weakly, that is, logarithmically, on the  metabolic and autoimmune costs. 

Our study is limited by the explicit assumption that the community is well-mixed: there is no geographic diversity and no migration.  In many naturally occurring environments, such as soils, the transport and mixing is slow, yet the cellular and viral co-evolution remains fast. That inevitably leads to the geographical separation (or spatial independence) of parallel evolutionary scenarios, which can produce distinct sets of defense and counter-defense systems. Thus, due to spatially-segregated evolution, the number of currently existing defense systems and evasion strategies can substantially exceed the simple estimates made using our models. We plan to consider the effects of spatial segregation and migration in the future.    

Another interesting effect we have noticed is that two strains of cells, with no defense systems and with the maximal sustainable number of such systems, often have the largest population densities. In the future, we plan to assess whether this effect is an artifact of our rather schematic simulation or is a real feature of the arms race. 

The implications of our findings could be quite general. Scenarios where two or more counterparts are engaged in antagonistic interactions that result in a buildup of defenses and counter-defenses abound in nature at all levels. In addition to the conflict between prokaryotic cells and their viruses analyzed here, such scenarios include interactions between various pathogens and immune systems of more complex organisms, cells that produce toxic substances such as antibiotics and cells that are sensitive to the action of these substences, and, more generally, all predator-prey type interactions. Furthermore, various social conflicts including crime and law enforcement, and even the actual arms race in between conflicting states also obey the  principles that we modeled in this work: each new defense level adds protection against a hostile action or a threat of the antagonistic counterpart while a counter-defense layer of the counterpart eliminates or reduces this protection. Since both defense and counter-defense carry costs and bearers of various numbers of layers are inevitably engaged in some form of competition, at some level it  becomes detrimental rather than beneficial to add more defense or counter-defense layers. It will be interesting to adapt our model to more specific biological and social scenarios and quantitatively probe these speculations. 

After completion of the main part of this study, we became aware of the publication that analyzes the dynamics of populations engaged in antagonistic interactions, albeit in a somewhat different way, \cite{kimchi2024bacterial}. It is assuring to see that that work essentially comes to conclusions similar to ours, even though several assumptions are quite different. 
%This brings up a provocative idea: whether such a limit on the number of layers a single species can sustainably hold translates into a limit on the total number of defense systems that can ever evolve? We hope to assess this connection in our future studies.  

\section {Acknowledgments}
AL acknowledges the support of the program "Prioritet-2030"  and would
like to thank the Department of Physics at the University of
Santiago of Chile for hospitality and the Moscow Institute of Physics and Technology for support.  
\vfill\eject

\bibliography{arms_race.bib}
\bibliographystyle{apalike}
\vfill \eject
\section* {Supplementary figures}
%\title{Sustaining the arms race: the maximal and optimal numbers of immune defense and counter-defense layers\\ Supplementary figures}

\renewcommand{\figurename}{Supplementary Figure}
\setcounter{figure}{0}
\begin{figure}[h!]
    \centering
   \includegraphics[width=.48\linewidth]{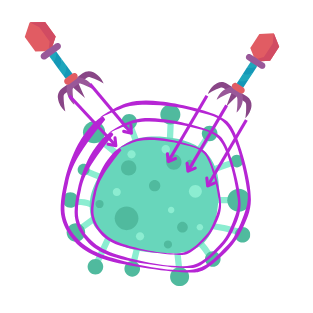}
   \includegraphics[width=.48\linewidth]{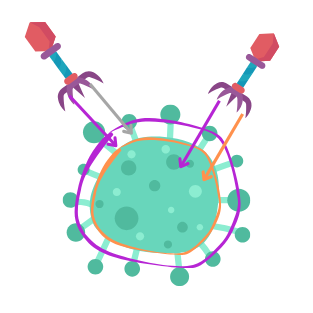} \\
    \caption{ Schemes of non-specific and specific immune and counter-immune systems are shown in the left and right panels. In the non-specific case, the attack rate of a virus with two counter-defense strategies (left) against a cell with three defense strategies is attenuated, while a virus with three counter-defense strategies (right) attacks the same cell without being inhibited by defense mechanisms. In the specific case, two distinct immune mechanisms, denoted by orange and purple shells, protect the cell from the attack by a virus (left) with one matching (purple) and one distinct (gray) counter-defense mechanisms. Yet the virus with both counter-defense mechanisms matching the corresponding defense strategies (right) overcomes the cell immunity.}
    \label{fs1}
\end{figure}

\begin{figure}[h!]
    \centering
   \includegraphics[width=.48\linewidth]{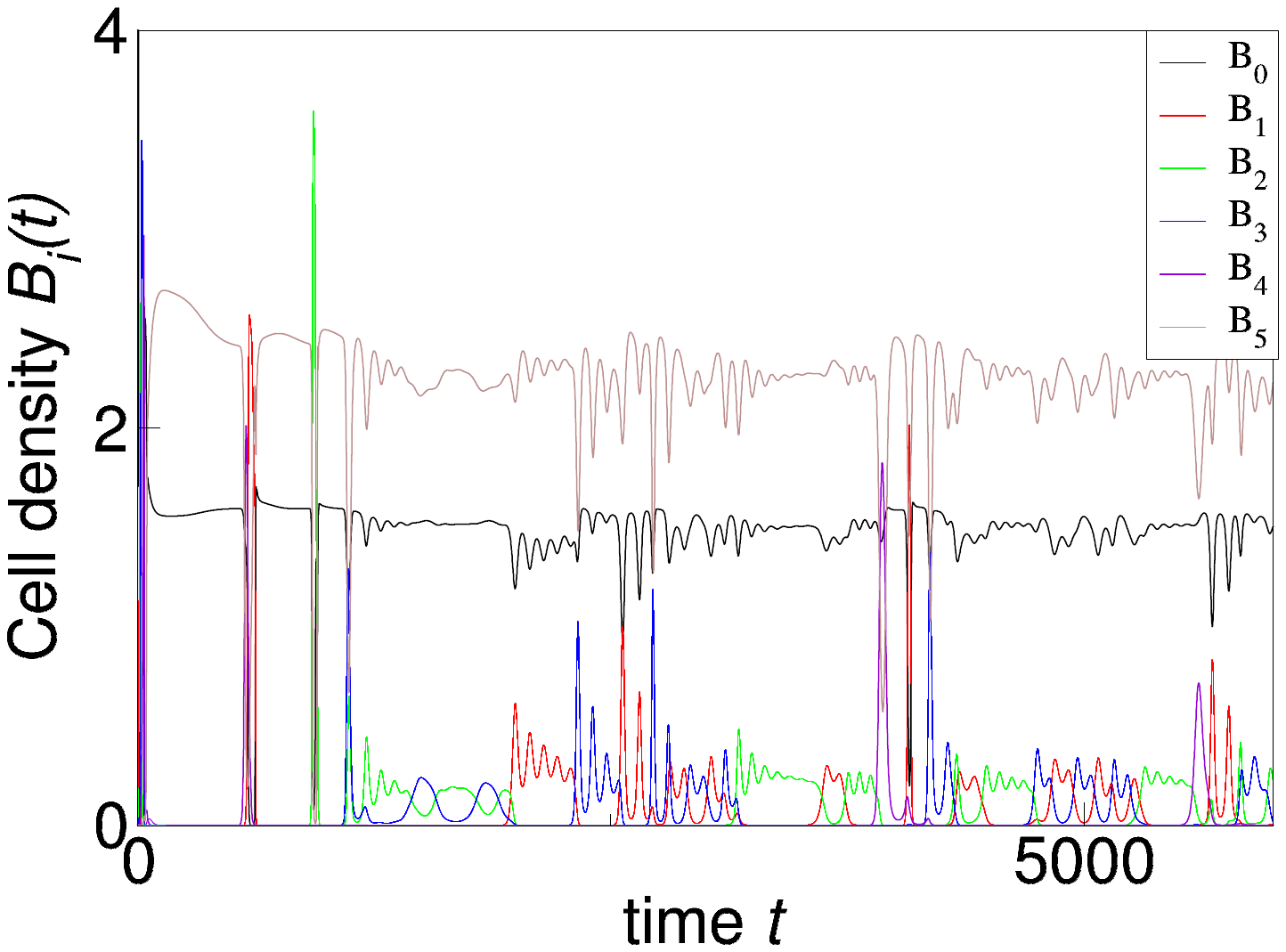}
   \includegraphics[width=.48\linewidth]{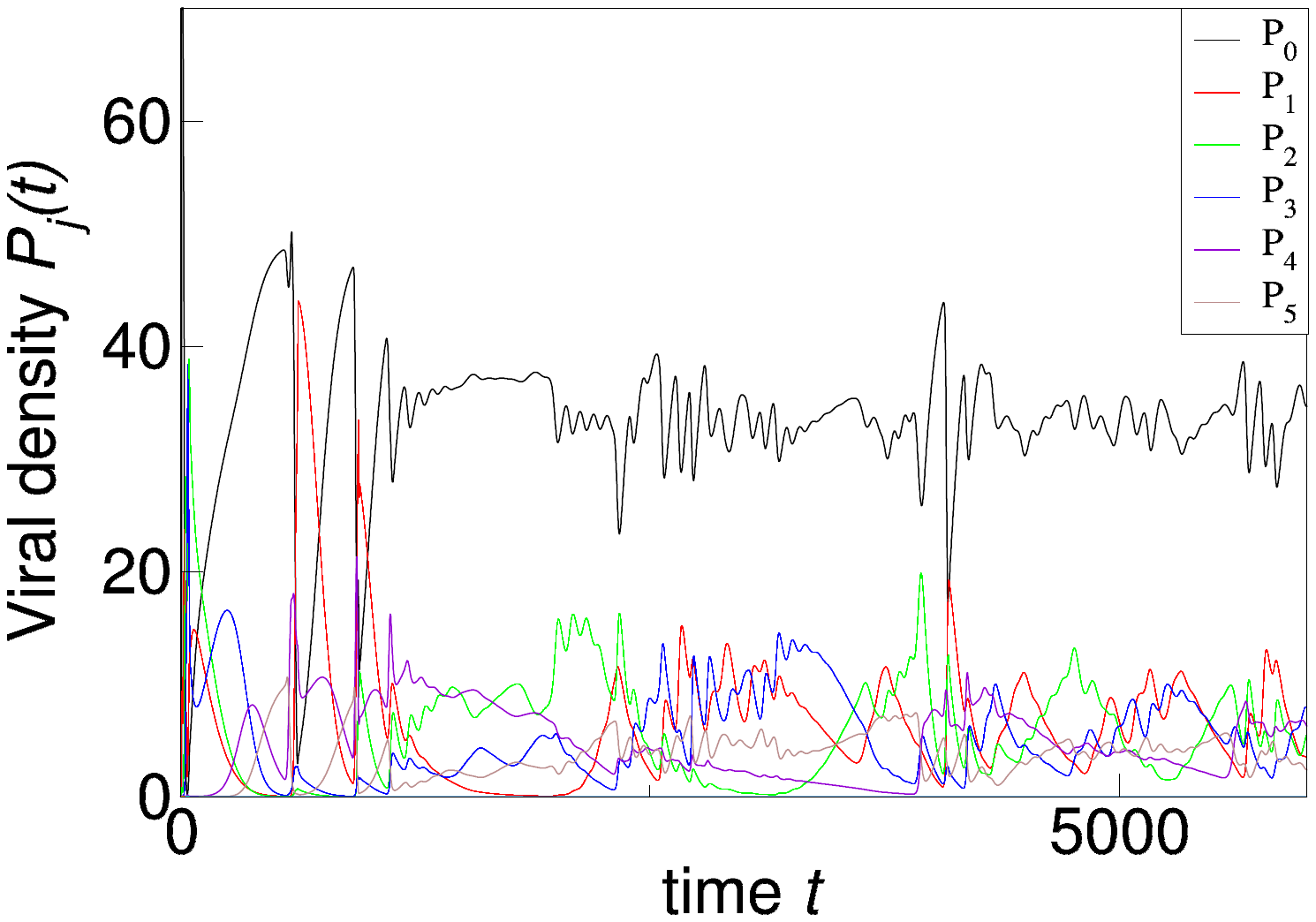} \\
   \includegraphics[width=.48\linewidth]{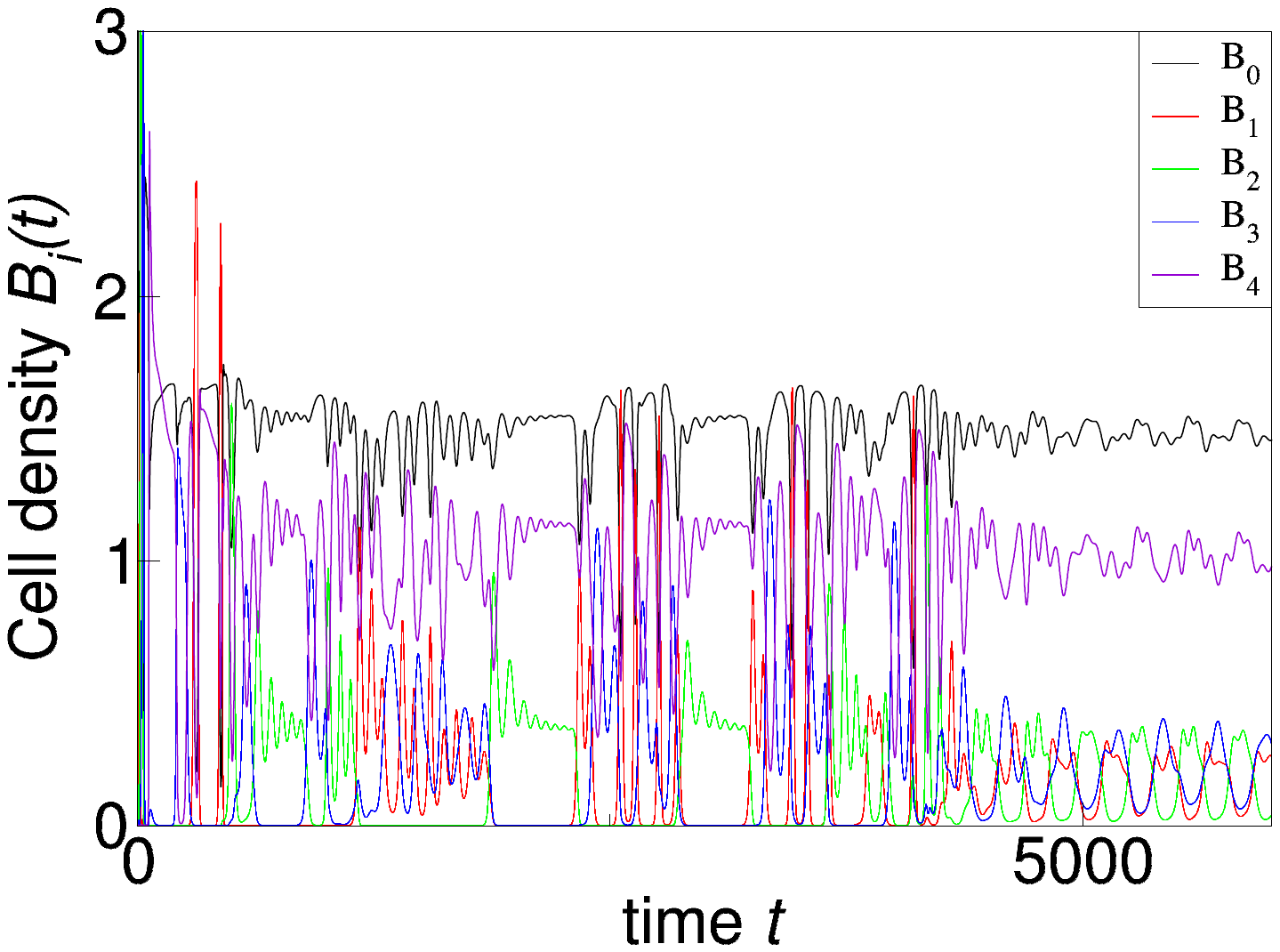}
   \includegraphics[width=.48\linewidth]{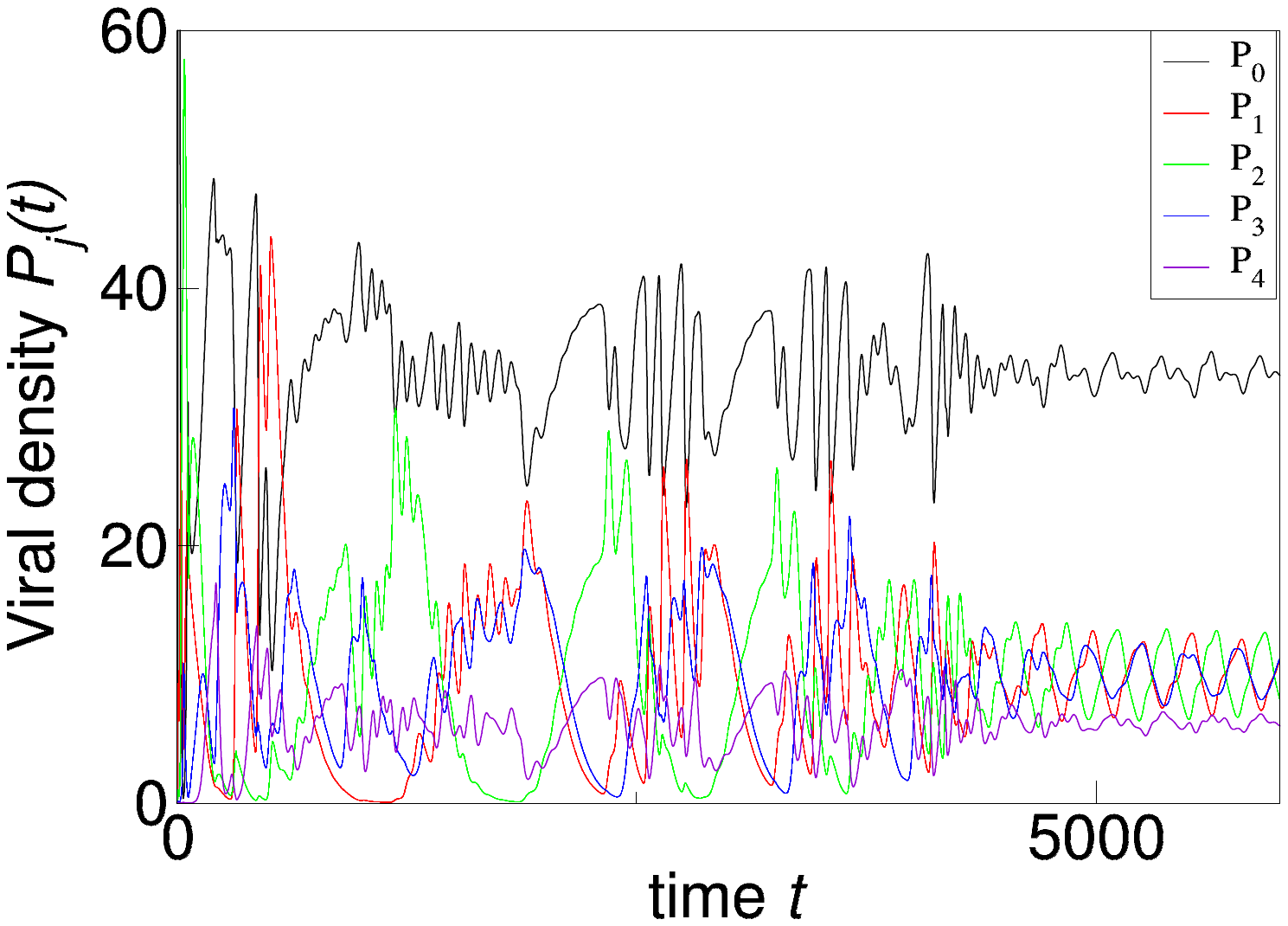} \\
    \caption{ Same as in  Fig.1 in the main text, but with the metabolic costs for cells being multiplicative and for viruses additive (given by Eqs. (3,4)) shown in the first row and the metabolic costs for cells being additive and for viruses multiplicative (given by Eqs. (2,5)) shown in the second row. 
    Six cellular (left panel) and six viral species (right panel) have non-vanishing population densities, so 5 defense and counter-defense layers are sustained.
   }
    \label{f3}
\end{figure}

\begin{figure}[h!]
    \centering
   \includegraphics[width=.48\linewidth]{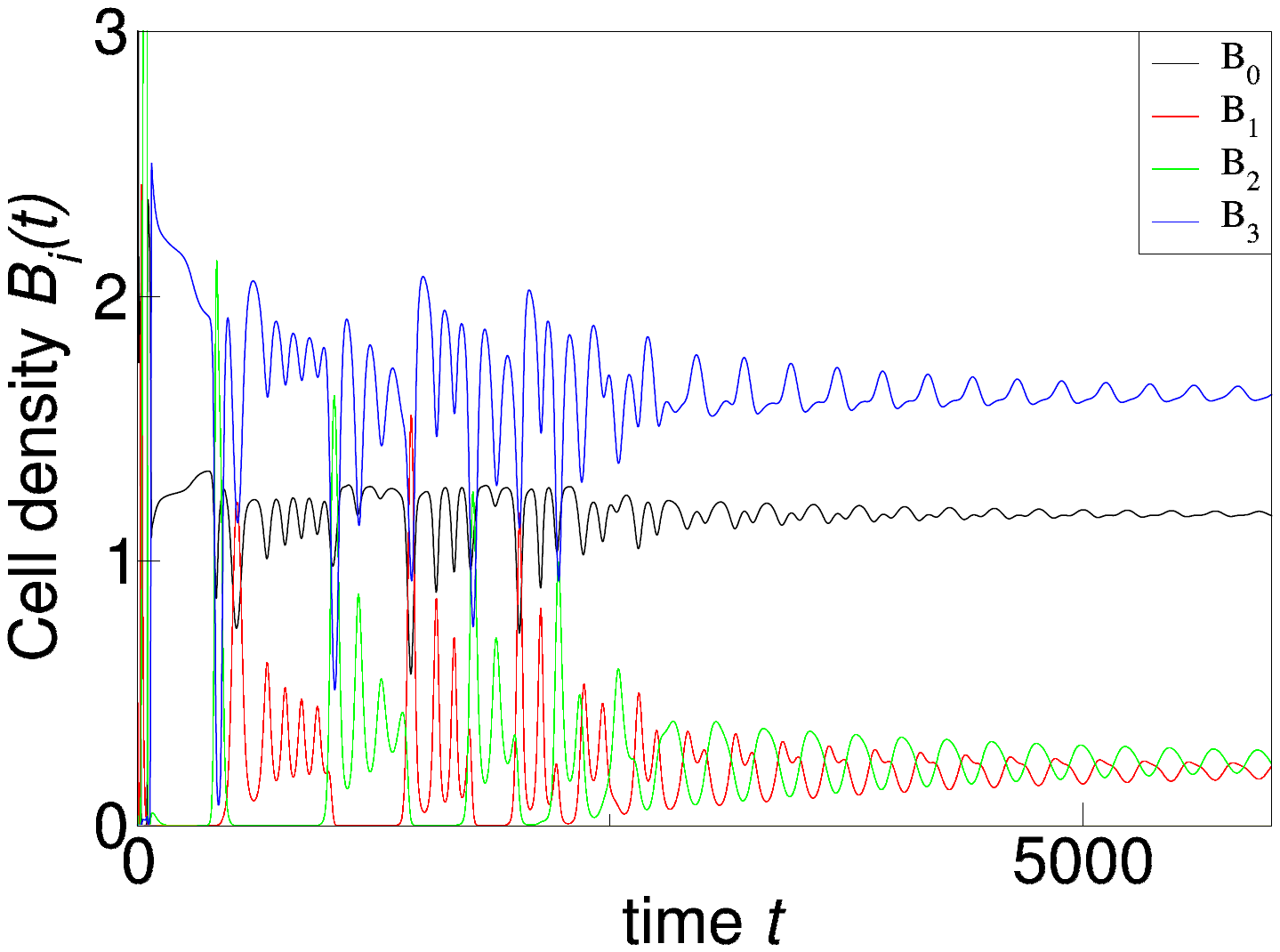}
   \includegraphics[width=.48\linewidth]{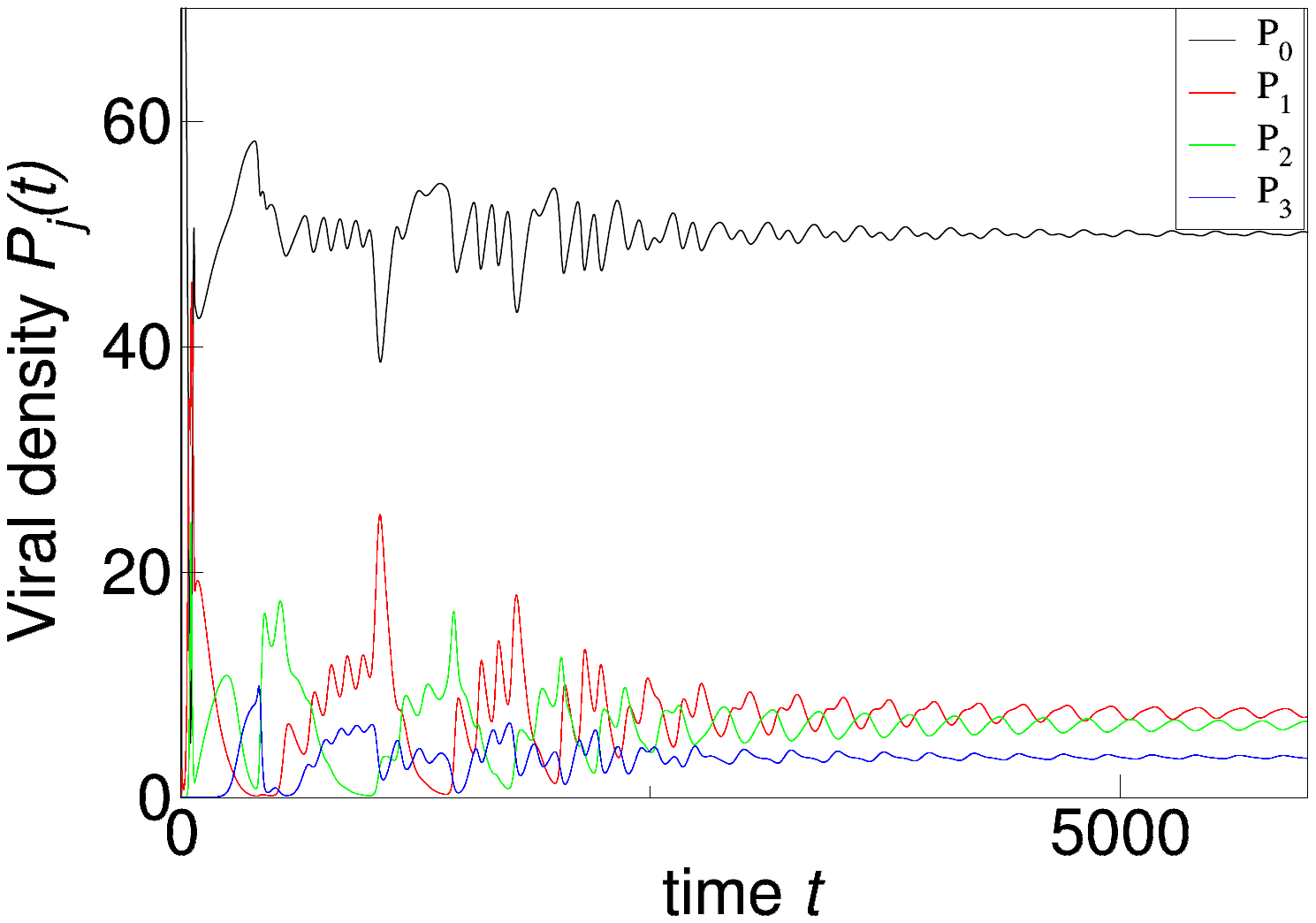} \\
    \caption{ Same as in top row of Fig.3 in the main text ($C_{\beta}=C_{\chi}=0.15$), but with the metabolic costs for cells and viruses being multiplicative instead of additive. Four cell (left panel) and  viral species (right panel) have non-vanishing population densities.}
    \label{f8}
\end{figure}

\begin{figure}[h!]
    \centering
   \includegraphics[width=.48\linewidth]{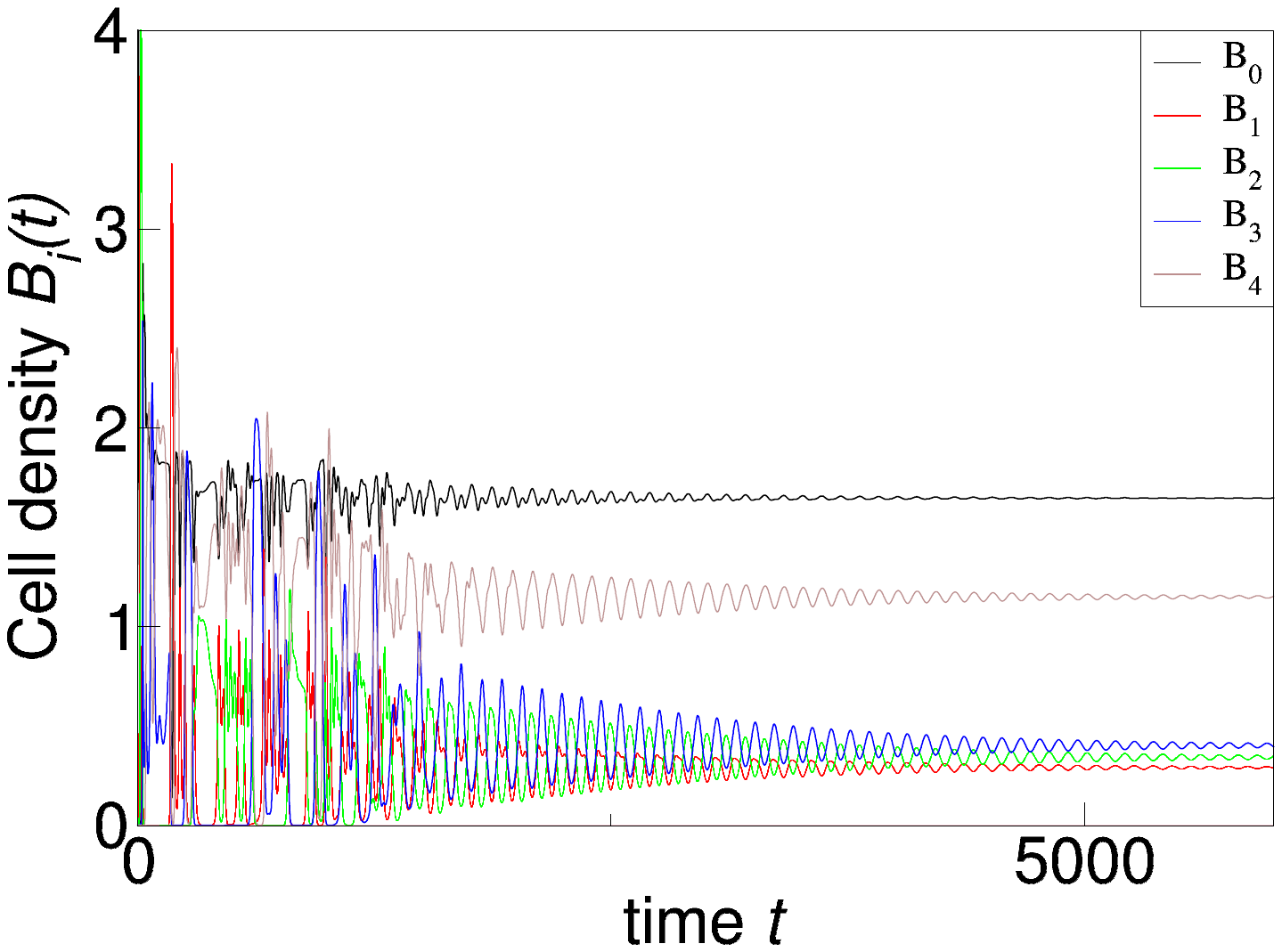}
   \includegraphics[width=.48\linewidth]{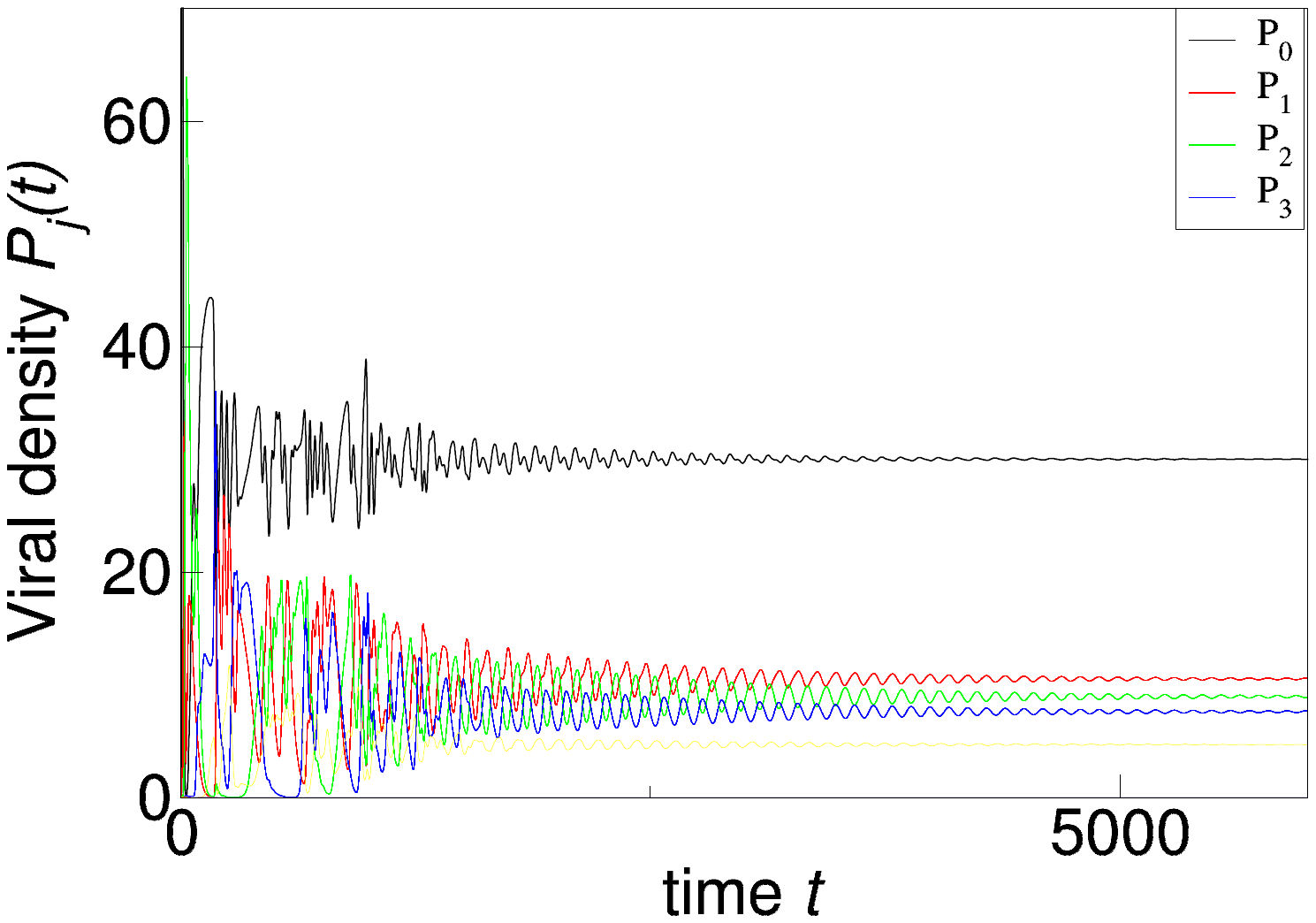} \\
    \caption{ Same as in Supplementary Fig. \ref{f8}, but with the reduction in attack rate $C_{\alpha}$ due to each additional level increased from $C_{\alpha}=0.3$ to $C_{\alpha}=0.5$. 
    This results in an increase in the number of defense layers from three to four (left panel), yet leaves the number of viral counter-defense systems limited by three (right panel).}
    \label{f10}
\end{figure}

\begin{figure}[h!]
    \centering
   \includegraphics[width=.48\linewidth]{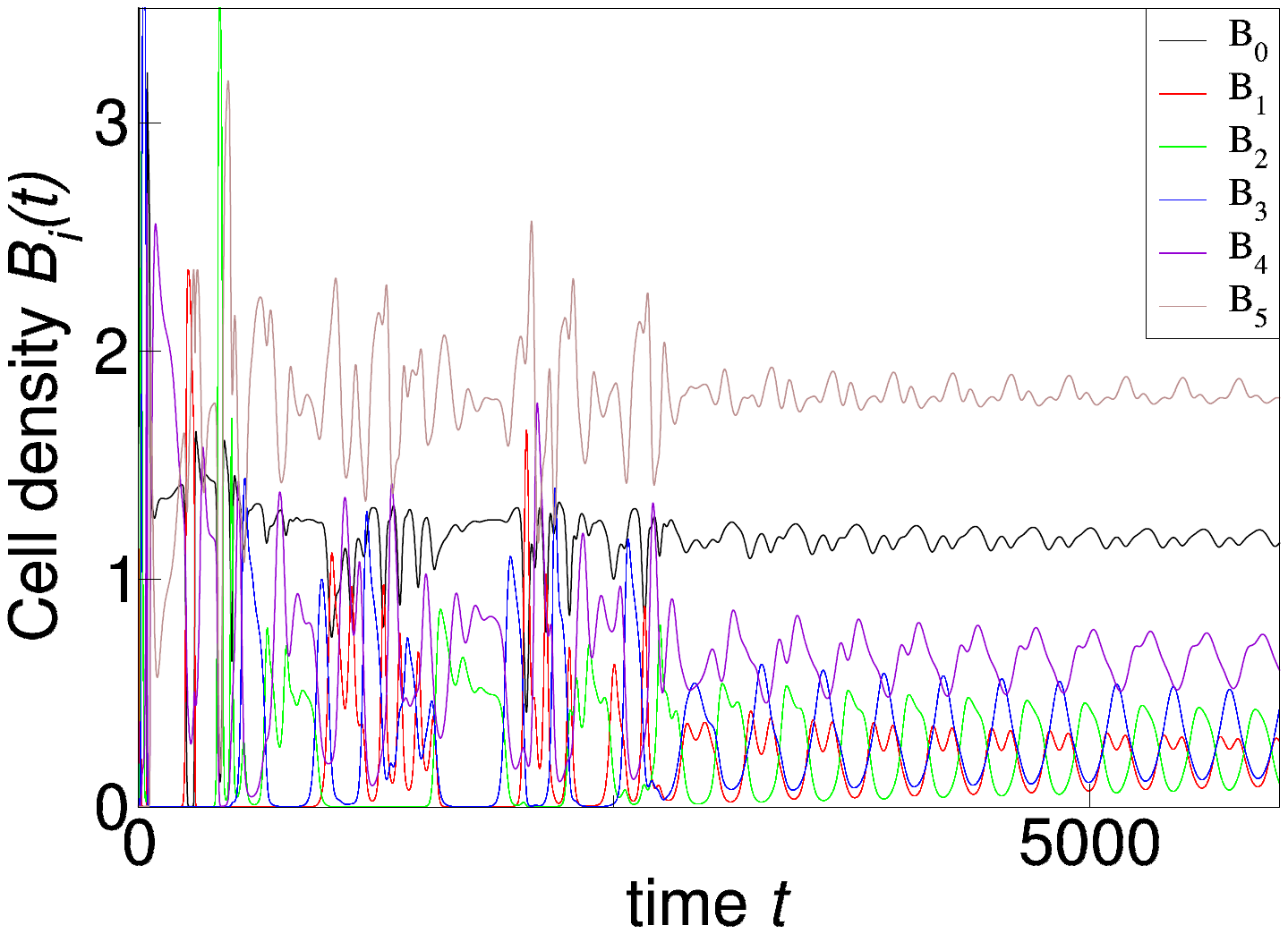}
   \includegraphics[width=.48\linewidth]{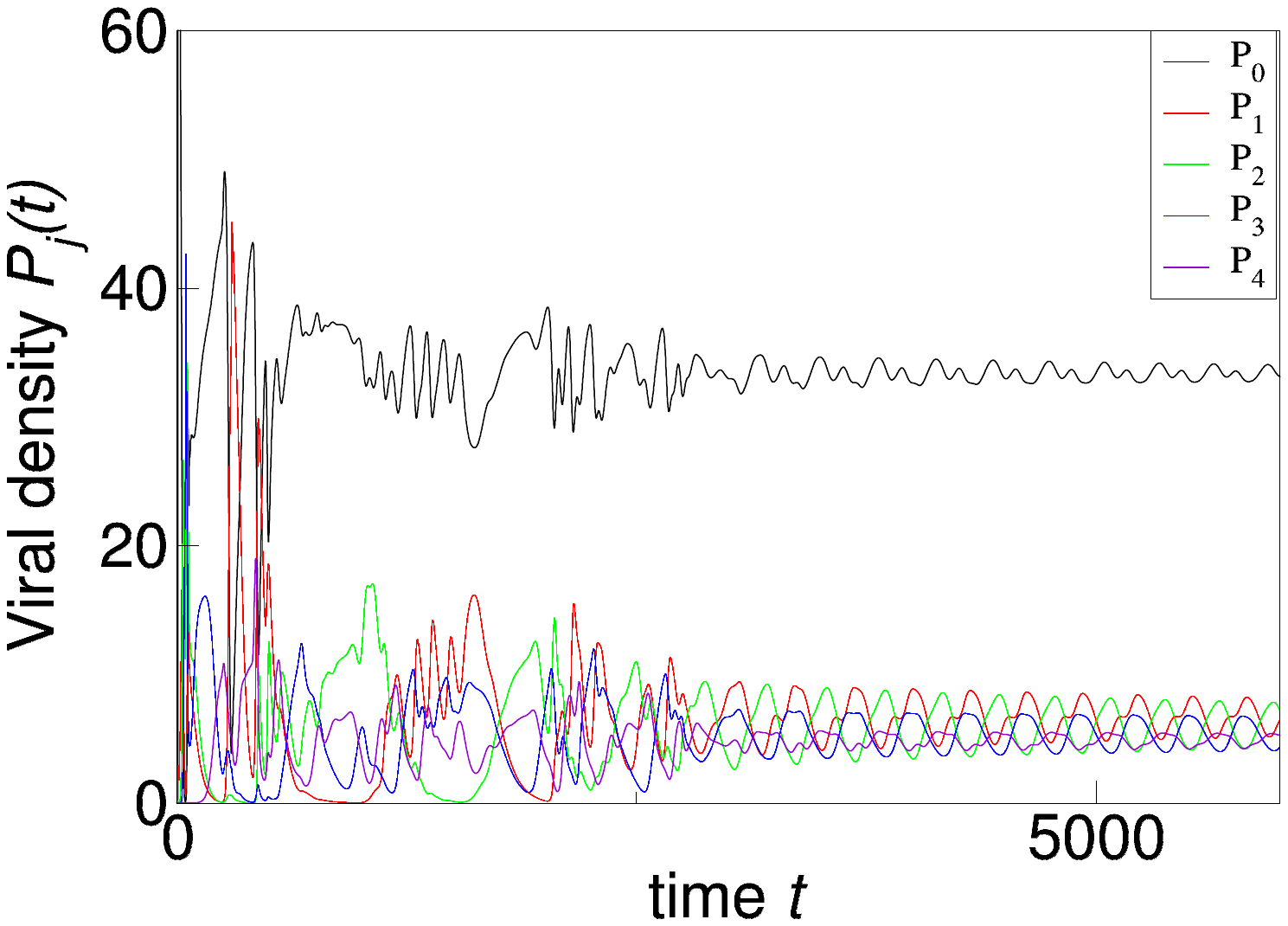} \\
    \caption{ Same as in Supplementary Fig. \ref{f8}, but with the metabolic costs for cells, $C_{\beta}=0.1$, being less than that for viruses,  $C_{\chi}=0.15$,
    As a  results, cells evolve 5 level of defense (left panel), yet viruses evolve only 4 evasion systems (right panel).}
    \label{f11}
\end{figure}
\begin{figure}[t!]
   \centering
   \includegraphics[width=.48\linewidth]{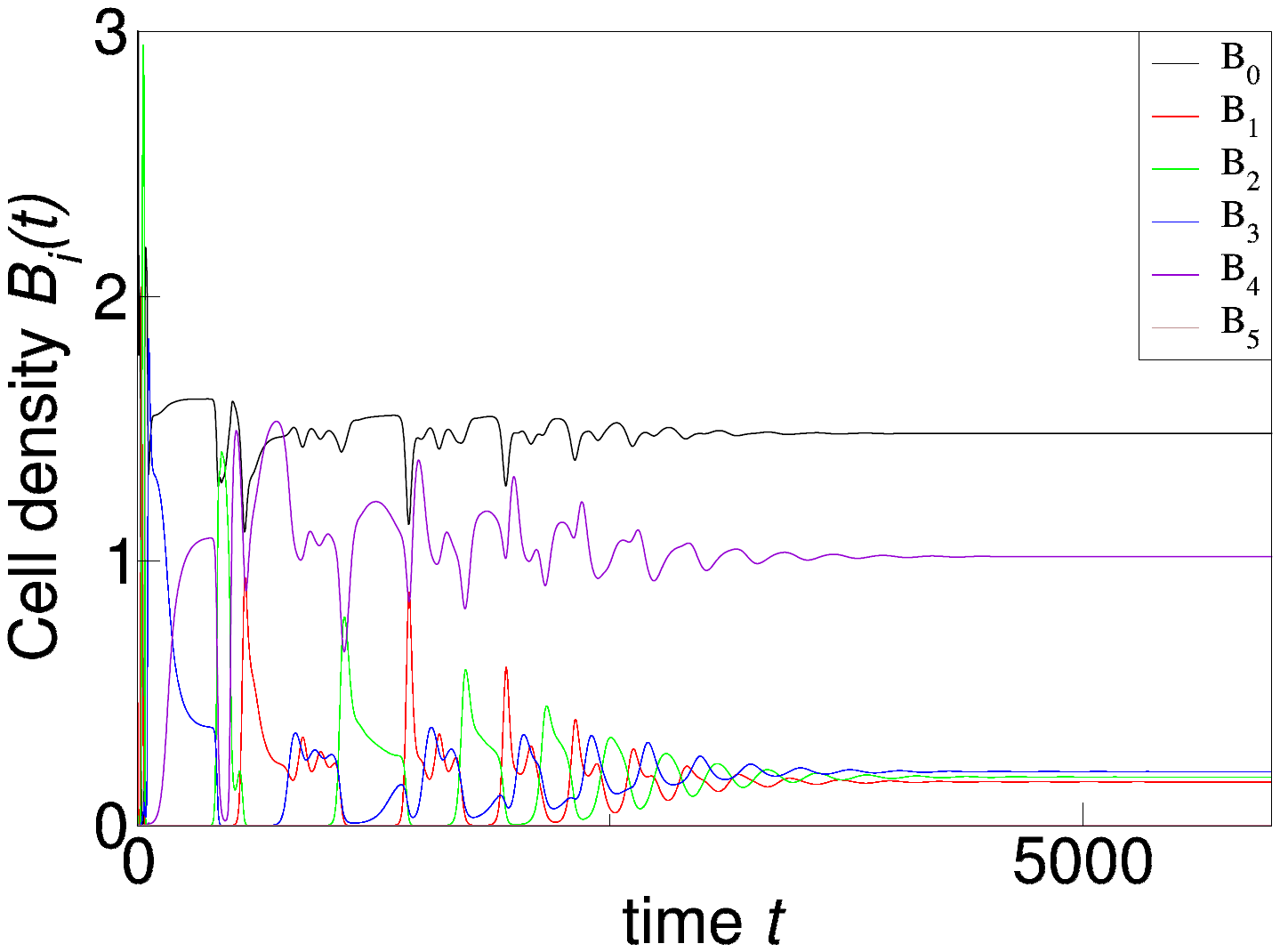}
   \includegraphics[width=.48\linewidth]{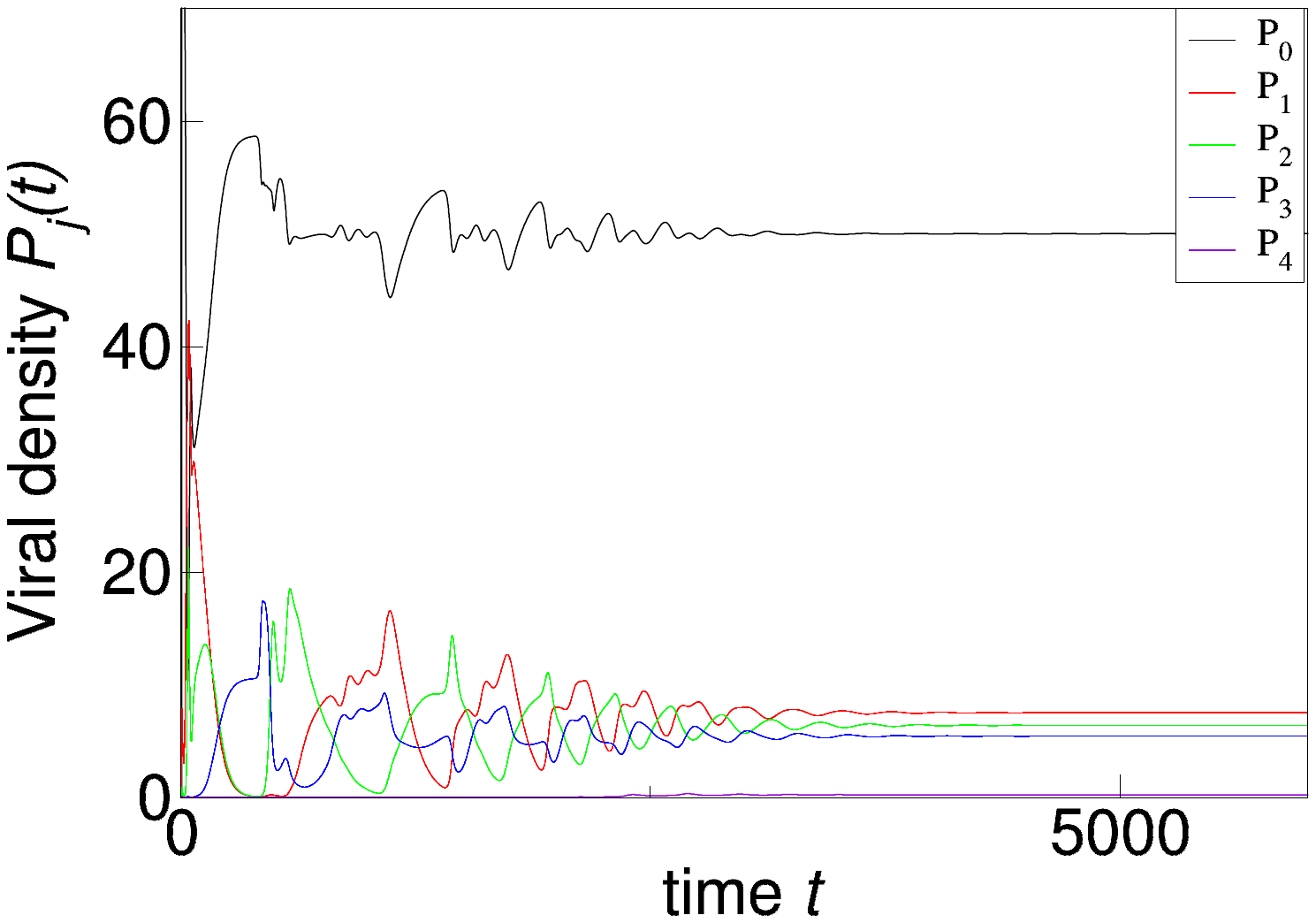} \\
    \caption{ Same as in second row in Supplementary Fig. \ref{f8}, but with the metabolic costs for cells, $C_{\beta}=0.15$, being greater than that for viruses,  $C_{\chi}=0.1$,
    As a  results, cells and viruses evolve 4 defense and counter-defense systems with the population of the viral strain with 4 layers being very small.}
    \label{f12}
\end{figure}

\end{document}